\documentstyle[prl,aps,epsfig]{revtex}         
\begin{document}
\draft               
\twocolumn[\hsize\textwidth\columnwidth\hsize\csname @twocolumnfalse\endcsname

\title{Incoherent optical switching of semiconductor resonator solitons}
\author {V.B.Taranenko, C.O.Weiss}
\address{Physikalisch-Technische Bundesanstalt 38116 Braunschweig/Germany}
\maketitle
\begin{abstract}
We demonstrate experimentally the bistable nature of the bright
spatial solitons in a semiconductor microresonator and show that
they can be created and destroyed by incoherent local optical
injection.
\end{abstract}
\pacs{PACS 42.65.Sf; 42.65.Tg; 42.70.Nq} \vskip1pc ]

Spatial optical solitons i.e. light beams propagating without
transverse spreading arise when diffraction is balanced by a
nonlinear process such as self-focussing in a nonlinear dispersive
or reactive medium. Light propagating inside an optical resonator
filled with a nonlinear medium can thus form stable filaments, or
localized structures (spatial solitons). These are free to move in
the resonator cross section (or move by themselves \cite{tag:1})
which implies their bistability, and ability to carry information.
The mobility of spatial solitons, however, makes them different
from arrangements of fixed binary elements, so that new types of
information processing have been considered, making use of spatial
resonator solitons.

Early realisations of such resonator solitons in slow materials
were given in \cite{tag:2,tag:3}. We investigated in the past
spatial resonator solitons of phase-type \cite{tag:4} and
intensity type \cite{tag:5}, including experiments demonstrating
large simultaneous collections of solitons \cite{tag:6} and their
manipulation \cite{tag:5} as is required for practical
applications. These experiments were conducted using slow
nonlinear materials for the sake of easy observeability of the
complex 2D space-time dynamics. For practical purposes, however,
speed is of prime importance and compatibility with semiconductor
technology is desirable. Spatial solitons and their switching in
semiconductor microresonators have therefore been predicted
theoretically recently \cite{tag:7,tag:9}. With the aim of
realizing spatial solitons in semiconductor resonators,
experiments were conducted recently, addressing passive resonators
\cite{tag:10} and resonators with population inversion
\cite{tag:11}. We showed the spontaneous formation of bright and
dark spatial solitons in \cite{tag:12}. We confirm here the
bistable nature of the bright spatial semiconductor resonator
solitons by the results of local switching experiments and
demonstrate the incoherent writing and erasing of the bright
solitons.

The experimental arrangement (FIG. 1) was essentially as described
in \cite{tag:10} and \cite{tag:12}. Light of a
Ti:Al$_{2}$O$_{3}$-laser around 855 nm wavelength illuminates the
semiconductor resonator sample. This consists of two Bragg mirrors
of about 99,5 $\%$ reflectivity and 18 pairs of
GaAs/GaAlAs-quantum-wells between them. The band edge and the
wavelength of the exciton line is at 849 nm. Observations are done
in reflection because the substrate material (GaAs) is opaque at
the working wavelength.

The laser light is modulated by a mechanical chopper to limit
illumination to durations of a few $\mu$s, in order to avoid
thermal nonlinear effects. The repetition rate of the
illuminations is 1 kHz, permitting stroboscopic recordings of the
dynamics or signal averaging. Part of the laser light is split
away from the main beam, with orthogonal polarisation, for local
injection into the illuminated sample area. The injection is
applied in pulses of several 10 ns duration using an
electro-optical modulator (EOM). The light reflected from the
sample is imaged onto a CCD camera. For time-resolved observations
the reflected light is passed through another EOM (50 ns aperture
time), which is opened with a variable delay with respect to the
start of the illumination. The 2D intensity can thus be recorded
for arbitrary moments during the illumination. Further, the
intensity in a particular point can be monitored by a small area
photodiode PD. In the observations reported, the intensity of the
main beam was chosen so that a bright soliton (dark in reflection)
would appear only at the center of the main beam which has
Gaussian intensity profile with a width of 30 $\mu$m. The small
area photodiode PD measures the intensity at this point.

\begin{figure}[htbf]
\epsfxsize=80mm \centerline{\epsfbox{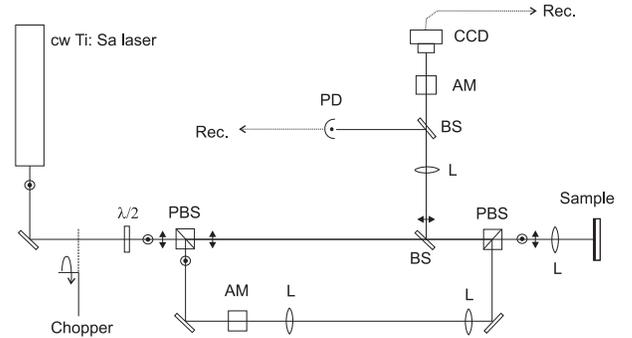}} \vspace{0.5cm}
\caption{Optical arrangement. ($\lambda$/2: halfwave plate, PBS:
polarizing beam splitters, AM: electro-optical amplitude
modulator, L: lenses, BS: beam splitters, PD: photodiode). Lenses
in the lower arm form a telescope, polarizations indicated.}
\end{figure}

FIG. 2 shows the switch-on of a bright soliton. The illuminating
intensity rises initially due to the mechanical chopper opening.
The maximum intensity is below the switching intensity for the
bistable resonator. At t $\approx$ 3.9 $\mu$s the injection beam
(orthogonal polarisation with respect to the illumination, width
12 $\mu$m) is opened for 70 ns to switch the resonator. During the
switch initiated by the injecting pulse a switching front travels
radially outward \cite{tag:10} and forms a switched area
surrounded by the switching front (dark in left inset in FIG. 2).
The switching area then collapses into a spot about 10 $\mu$m
diameter (see right inset in FIG. 2), the expected size of a
soliton for this resonator (details see \cite{tag:12}). This
collapse takes place from 4 to 5 $\mu$s. After 5 $\mu$s a
stationary soliton exists, recorded in the right inset of FIG. 2.
When the incident illumination (dotted trace) is finally decreased
(chopper closes) the soliton switches off.

\begin{figure}[htbf]
\epsfxsize=70mm \centerline{\epsfbox{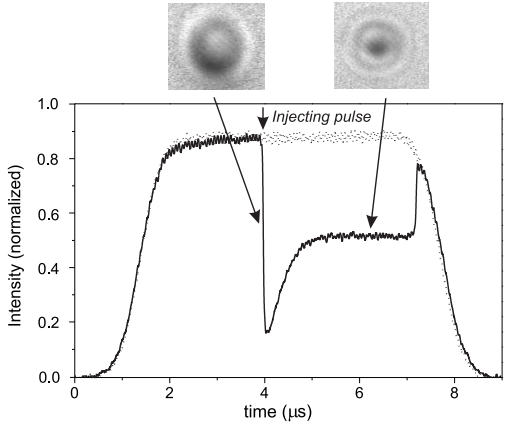}} \vspace{0.5cm}
\caption{Recording of the intensity reflected from the sample at
the center of the illuminating beam, when switching a soliton on.
Snapshot pictures show circular outward travelling switching front
(left) and a soliton (right). Dotted trace: incident intensity.\\
The intensity value of 1 is the same absolute intensity in FIG. 2,
3, 4. }
\end{figure}

Although we do not understand presently the mechanism by which the
relatively slow collapse of the switching front occurs, FIG. 2
demonstrates that the bright soliton can be switched on by an
external control beam, implying the bistability of the soliton.
The switch-on of the soliton would presumably proceed more
directly if the injection beam size, intensity, phase and
polarization were matched to the final soliton dimensions.

FIG. 3 shows conversely the switching off of a bright soliton
(dark in reflection). The illumination intensity in FIG. 3 is
chosen above the switching intensity of the resonator so that a
soliton forms spontaneously as described in \cite{tag:12} during
the transient phase from about 2.4 to 3.5 $\mu$s. At about 3.5
$\mu$s a stationary soliton is existing (see central inset in FIG.
3). At about 3.9 $\mu$s the injection beam (same properties and
alignment as for Fig. 2) is opened. This switches the soliton off
and returns the whole resonator to the unswitched state (see the
right inset in FIG. 3). We mention that the state after the
switch-back is stable here, although at its intensity it was
unstable in the beginning (t $\approx$ 2.4 $\mu$s). Measurements
showed that after the switch-off the threshold of instability was
a few percent higher than initially. Raising the background
intensity slightly led to renewed spontaneous appearance of the
soliton with the pronounced feature of critical slowing. One might
tentatively ascribe the small increase of the instability
threshold to heating of the material during the time of formation
and existence of the soliton, during which the intensity and
dissipation in the resonator is high.

\begin{figure}[htbf]
\epsfxsize=70mm \centerline{\epsfbox{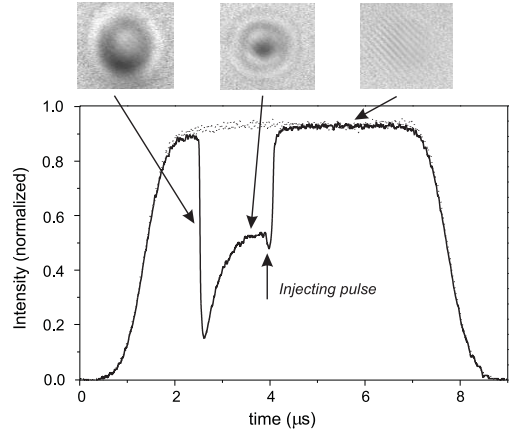}} \vspace{0.5cm}
\caption{Recording of the intensity reflected from the sample at
the center of the illuminating beam. A soliton is formed
spontaneously and then switched off by external injection. The
insets left, center, right show intensity snapshots, namely
switching front (as FIG. 2), soliton (as FIG. 2) and unswitched
state, resp. Dotted trace: incident intensity. }
\end{figure}

FIG. 4 shows that the switching off of the soliton requires a
minimum intensity in the external beam. Here the external beam is
opened at t $\approx$ 3.9 $\mu$s with an intensity 10 $\%$ smaller
than in FIG. 3. The soliton in this case is transiently perturbed,
but remains stably switched-on.

\begin{figure}[htbf]
\epsfxsize=70mm \centerline{\epsfbox{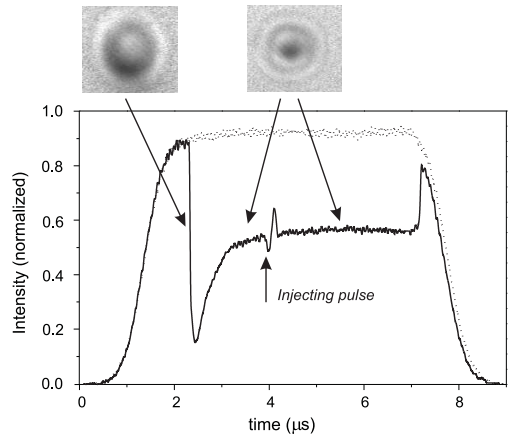}} \vspace{0.5cm}
\caption{Failed switch-off of soliton.\\ Intensity of the injected
beam too small to switch soliton off. Dotted trace: incident
intensity.}
\end{figure}

In summary, we have shown that bright spatial solitons of a
semiconductor resonator can be switched on and off by an external
incoherent address beam. Thus we demonstrate that such solitons
are controllable as required for applications. The bistable nature
of the solitons is unambiguously demonstrated. The switch-on
mechanism observed presently is too slow for fast processing
applications. We attribute this to an insufficient match of the
injection beam field with the soliton and suppose that a matched
injection beam should directly switch on the bright solitons,
without the long transient soliton formation phase. The details of
the incoherent switching mechanism observed here will be clarified
in the near future.\\

Acknowledgement

This work was supported by ESPRIT LTR project PIANOS. The
quantum-well semiconductor sample was provided by R.Kuszelewicz,
CNET, Bagneux, France.

\end{document}